\documentclass[journal]{IEEEtran}
\IEEEoverridecommandlockouts

\usepackage{cite, hyperref}
\usepackage{amsmath,amssymb,amsfonts,amsthm}
\usepackage{algorithmic}
\usepackage{graphicx}
\usepackage{textcomp}
\usepackage{xcolor}
\usepackage{float}
\usepackage{multirow}

\def\BibTeX{{\rm B\kern-.05em{\sc i\kern-.025em b}\kern-.08em
    T\kern-.1667em\lower.7ex\hbox{E}\kern-.125emX}}

\newcommand{\M}[4]{
\renewcommand{\arraystretch}{0.8}
\setlength{\arraycolsep}{2pt}
\fontsize{7}{12}\selectfont
\begin{bmatrix}#1 & #2\\ #3 & #4\end{bmatrix}}
\newcommand{\vT}[2]{\left[#1,\,#2\right]^{\mathsf{T}}}
\newcommand{\bfg}[1]{\boldsymbol{#1}}
\newcommand{\ul}[1]{\underline{#1}}
\newcommand{\ull}[1]{\underline{\underline{#1}}}

\newcommand{\Iden}{\ull{\boldsymbol{I}}}

\newcommand{\mZmat}{\ull{\bfg Z}}
\newcommand{\mZmateq}{\ull{\bfg Z_{\rm eq}}}
\newcommand{\mZmateqp}{\ull{\bfg Z_{\rm eq}'}}
\newcommand{\mZmateqpp}{\ull{\bfg Z_{\rm eq}''}}

\newcommand{\uletavecam}{\widetilde{\ull{\bfg \eta}}'}
\newcommand{\uletaavecam}{\widetilde{\ull{\bfg \eta}}''}

\newcommand{\szero}{\ull{\bfg S}}
\newcommand{\sone}{\ull{\bfg S}'}
\newcommand{\stwo}{\ull{\bfg S}''}

\newcommand{\azero}{}
\newcommand{\aone}{}
\newcommand{\atwo}{}
\newcommand{\bone}{}
\newcommand{\btwo}{}
\newcommand{\ctwo}{}

\begin{document}
\bstctlcite{BSTcontrol}

\title{Practical Framework for Power System Strength}

\author{Ignacio Ponce,~\IEEEmembership{Graduate Student Member,~IEEE,}
        and Federico Milano,~\IEEEmembership{Fellow,~IEEE}%
\thanks{I.  Ponce and F.  Milano are with the School of Electrical and Electronic Engineering, University College Dublin, Belfield Campus, D04V1W8, Ireland.
e-mails: ignacio.poncearancibia@ucdconnect.ie, federico.milano@ucd.ie}%
\thanks{This work is supported by the Sustainable Energy Authority of Ireland (SEAI) by funding I.  Ponce and F.  Milano under project FRESLIPS, Grant No.~RDD/00681.}%
\vspace{-7mm}
}

\maketitle

\begin{abstract}
A novel analytical framework for power system strength was recently introduced in the IEEE Transactions on Power Systems, providing a unified formulation for assessing voltage and frequency strength.  Building upon this formulation, this paper addresses a series of practical challenges for translating the theoretical framework into a real-world application.  In particular, simplified analytical solutions for network-wide bus-level strength metrics are provided, together with compact expressions to capture the impact of relevant devices on strength.  In addition, novel normalized strength metrics at a device level are defined, enabling the comparison of strength across different systems.  A strength source model is introduced to study the behavior of devices under varying strength conditions.  Finally, the framework is implemented in a real-world study case, demonstrating its applicability and potential as a practical tool for a comprehensive strength assessment.  
 
\end{abstract}

\section{Introduction}

% \subsection{Motivations and Literature Review}

The concept of \textit{power system strength} has gained increased attention in the last years due to its importance for ensuring an adequate operation of power systems with high shares of inverter-based resources (IBRs).  Strength is commonly understood as the ability of a system to resist changes in the three-phase AC voltage at a certain location, typically reflected through variables such as the voltage magnitude, phase angle, and frequency \cite{cigrestrength}.  While traditional metrics such as the short-circuit level (SCL) or inertia have been widely used in practice, they have fallen short for quantifying strength in power systems with an increasingly heterogeneous mix of devices \cite{mancarella,damanik,qays}.  More advanced strength metrics and novel formulations have recently emerged to account for these changes, e.g., \cite{iscr,sdscr,gsimp,task2018force,kim2020,gscr,imr,sanni} focused on voltage magnitude strength, and \cite{zeng, gosh,pinheiro, pinheiro2, idenareas} focused on frequency strength.  However, their applicability is often limited to specific devices, operating conditions, and time scales.  Besides, they often remain dependent on strong approximations, and still rely on a completely decoupled assessment of voltage magnitude strength and frequency strength, even though they have shown increasingly coupled dynamics \cite{taskforce2, cong, bernal}.  Given that both variables are ultimately properties of the same three-phase AC voltage waveform, this motivates the development of a unified strength assessment framework.

In a recent paper \cite{strength1}, we have introduced a novel analytical framework for power system strength based on the complex frequency concept \cite{ComplexFreq}.  The formulation is composed of twelve indicators grouped in three dynamical orders that quantify the sensitivity of the three-phase AC voltage to perturbations.  This formulation establishes a rigorous theoretical foundation for strength assessment, suitable for data-driven or model-driven approaches.  In \cite{strength1}, exact model-driven analytical solutions are provided for relevant power system devices, along with a systematic procedure for extending the strength assessment to arbitrary device models and time scales.

Several challenges remain in translating the proposed framework into a tool suitable for practical power system strength analysis, planning and operation.  In particular, the exactness of the formulation proposed in \cite{strength1} comes at the cost of increased complexity, requiring detailed knowledge on the system and perturbation that are hardly available in practice.  Moreover, key aspects for a real-world application---such as simplified analytical expressions, typical ranges for acceptable strength values, and normalized device-level metrics---were not included in the theoretical work due to scope alignment and space constraints.

\subsection{Contributions}

This paper addresses these challenges by developing a practical implementation of the analytical framework for power system strength.  The goal is to progress on bridging the gap between the theoretical foundation and a real-world engineering application.  The main contributions of the paper are as follows:
\begin{itemize}
    \item A simplified analytical framework for power system strength that removes the dependency on post-disturbance variables and reduces computational complexity.  
    \item The definition of normalized device-level strength metrics that allow comparing strength levels across systems and operating conditions.
    \item The dynamic model of a novel ``strength source'' that represents an external equivalent network with predefined strength levels.
    \item Guidelines for defining acceptable levels of strength based on the proposed formulation.
\end{itemize}

\subsection{Organization}

The remainder of the paper is organized as follows.  Section \ref{sec:notation} introduces the notation used in the paper.  Section \ref{sec:framework} presents our strength framework, including relevant background, the core formulation, simplified solutions and analytical expressions for evaluating strength.  Section \ref{sec:dynamic} presents a dynamic analysis that provides guidelines for defining acceptable strength levels, and introduces a novel model of strength source.  Section \ref{sec:studycase} describes the implementation of our framework in a real study case.  Finally, Section \ref{sec:conclusion} concludes the paper by providing general remarks, recommendations, and outlines future work.
% \newpage

\section{Notation}
\label{sec:notation}

A bar over a symbol indicates a complex quantity:
\begin{equation}\label{eq:not1}
\bar{v}\in\mathbb{C},\quad\bar{v}=v_{\rm d}+\jmath\, v_{\rm q}\,,
\end{equation}
where $v_{\rm d},v_{\rm q}\in \mathbb{R}$ and $\jmath$ is the imaginary number.

A double bar below a symbol indicates a $2\times2$ square matrix:
\begin{equation}\label{eq:not2}
\ull{v}\in\mathbb{R}^{2\times2},\quad \ull{v}=
\begin{bmatrix}
v_{11} & v_{12} \\
v_{21} & v_{22}
\end{bmatrix} ,
\end{equation}
where $v_{11},v_{12},v_{21},v_{22}\in \mathbb{R}$.

A single bar below a symbol denotes a $2\times1$ vector:
\begin{equation}\label{eq:not3}
    \ul{v}\in\mathbb{R}^{2\times1}=
    \begin{bmatrix}
    v_{\rm d}\\
    v_{\rm q}
    \end{bmatrix}=\vT{v_{\rm d}}{v_{\rm q}} ,
\end{equation}
where $v_{\rm d},v_{\rm q}\in \mathbb{R}$ and $\mathsf{T}$ is the transpose operator.

Bold lowercase letters (e.g., $\bfg v$) denote column vectors containing variables for every bus of the network, whereas bold uppercase letters (e.g., $ \bfg Y$) denote square matrices containing variables that relate every pair of buses of the network.  Whenever bars are used above or below bold letters (e.g., $\ull{\bfg Z}$), it indicates that each of their elements are of the corresponding type according to (\ref{eq:not1})-(\ref{eq:not3}).  

A dot over a scalar quantity denotes its time derivative:
\begin{equation}
    \dot{x}=\frac{dx}{dt} \,.
\end{equation}

Regarding the non-scalar objects given by (\ref{eq:not1})-(\ref{eq:not3}), a dot over the object also denotes its time derivative.  In case the reference frame of the quantity is in motion, e.g., $\bar{v}=v_{\rm d}+\jmath\,v_{\rm q}$ in rotating $\rm dq$ coordinates, the dot over the dynamic object denotes its time derivative relative to a common stationary reference frame rotating at the fundamental frequency $\omega_0$.  Therefore, $\dot{\bar{v}}=\dot{v}_{\rm d}+\jmath\,\dot{v}_{\rm q}+\,\jmath(\omega_{\rm dq}-\omega_0)\bar{v}$, where $\omega_{\rm dq}$ is the angular speed of the rotating $\rm dq$ coordinates of $\bar{v}$.

Finally, the symbol $\Delta$ denotes a finite difference operator, so-called the Delta operator in \cite{strength1}, whereas the symbol $\sim$ over a quantity denotes the \textit{instantaneous arithmetic mean} as defined in the addendum of \cite{strength1}.  For instance, $\Delta x=x^{+}-x^{-}$ and $\widetilde{x}=0.5\,(x^{+}+x^{-})$, where $x^+$ and $x^-$ denote the post- and pre-disturbance values of $x$, respectively.

\section{Analytical framework}
\label{sec:framework}

In this section we introduce our analytical framework for evaluating power system strength.

\subsection{Background}

In this work, strength is conceived as a property of the voltage at a bus.  It quantifies the ability of the system to resist changes in the voltage, particularly to its magnitude, phase, as well as its frequency and rate of change of frequency.  Without lack of generality, voltage strength is evaluated with respect to changes in the current injected at the bus.

The concept of strength is formalized using the concept of the complex frequency \cite{ComplexFreq}.  Consider a three-phase voltage in balanced conditions represented as a dynamic Clarke vector $\bar{v}\in \mathbb{C}\,|\,\bar{v}=v\cos\theta+\jmath\, v\sin\theta$.  The first-order relative complex frequency of the voltage is denoted as $\bar{\eta}'$ and defined as:
% \footnote{Prime superscripts must not be confused with the time derivative operator.  Even though for the first- and second-order complex frequencies it holds that $\bar{\eta}''=\dot{\bar{\eta}}'$, this is coincidental.  Prime superscripts are simply graphical signs that indicate the strength order, in the same way subscripts $\rm dq$ indicate a d-axis or q-axis variable.}:
%
\begin{equation}
    \bar{\eta}'=\rho+\jmath\,(\omega-\omega_0)=\frac{\dot{v}}{v}+\jmath\, (\dot{\theta}-\omega_0),\quad v\neq0\,,
\end{equation}
whereas the second-order relative complex frequency is denoted as $\bar{\eta}''$ and defined as:
\begin{equation}
    \bar{\eta}''=\sigma+\jmath\,\gamma=\dot{\rho}+\jmath\,\dot{\omega}\,.
\end{equation}

%For the developments of the framework, the following notation is used:
%\begin{itemize}
%    \item A single bar under a quantity denotes a $2\times1$ vector, e.g.,  $\ul{z}=\left[a\quad b\right]^{\mathsf T}$.
 %   \item Double bars under a quantity denotes a $2\times2$ matrix, e.g., $\ull{Z}=\M{a}{b}{c}{d}$.
%\end{itemize}
\subsection{Formulation}
The definition of strength is formalized through the formulation presented in \cite{strength1}.  It comprises three categories of strength indicators.  Zero-order indicators characterize the stiffness of the voltage magnitude and phase angle, first-order indicators describe the stiffness of the complex frequency, and second-order indicators capture the stiffness of the rate of change of the complex frequency.  The three equations below implicitly define the zero-, first-, and second-order strength indicators, respectively denoted as $\ull{S}$, $\ull{S}'$ and $\ull{S}''$:
\begin{align}
        \left[\begin{array}{cc}\Delta v/v\\\tan(\Delta\theta)\end{array}\right]&=
        \ull{S}\cdot
        \left[\begin{array}{cc}\Delta \imath_{\rm p}\\\Delta \imath_{\rm q}\end{array}\right] , \label{eq:f0}\\
        \left[\begin{array}{cc}\Delta \rho\\\Delta \omega\end{array}\right]&=
        \ull{S}'\cdot
        \left[\begin{array}{cc}\Delta \imath_{\rm p}\\\Delta \imath_{\rm q}\end{array}\right] , \\
        \left[\begin{array}{cc}\Delta \sigma\\\Delta \gamma\end{array}\right]&=
        \ull{S}''\cdot
        \left[\begin{array}{cc}\Delta \imath_{\rm p}\\\Delta \imath_{\rm q}\end{array}\right] , \label{eq:f2}
\end{align} 
where:
\begin{align}
    \ull{S}&=\M{S_{v\imath_{\rm p}}}{S_{v\imath_{\rm q}}}{S_{\theta\imath_{\rm p}}}{S_{\theta\imath_{\rm p}}},\, \ull{S}'=\M{S_{\rho\imath_{\rm p}}}{S_{\rho\imath_{\rm q}}}{S_{\omega\imath_{\rm p}}}{S_{\omega\imath_{\rm p}}},\,\ull{S}''=\M{S_{\sigma\imath_{\rm p}}}{S_{\sigma\imath_{\rm q}}}{S_{\gamma\imath_{\rm p}}}{S_{\gamma\imath_{\rm p}}}\,,
\end{align}
and $\imath_{\rm p}$ and $\imath_{\rm q}$ are the active and reactive currents.

Each category comprises four scalar indicators, resulting in twelve metrics in total.  Together, these twelve quantities characterize the voltage's strength at a certain location.  There are two alternatives to calculate this set of metrics: a data-driven approach based on measurements or simulation results, or, an analytical approach based on a mathematical model of the system.  In \cite{strength1}, we adopted an analytical approach that led to explicit solutions for the strength metrics as a function of the system's variables and parameters.  The system representation used therein minimized modeling assumptions which, helps obtaining exact solutions.  However, such exactness comes with great requirements in terms of information, such as knowing the pre- and post-disturbance values of algebraic variables of the system.  In the remainder of this section we provide a simplified and more practical version of the solutions and analytical expressions presented in our previous work.  

\subsection{Simplified analytical solution}\label{subsec:simple}
The exact analytical solutions for the strength metrics of the network found in \cite{strength1} are:
\begin{align}
    \szero &= \widetilde{\bfg v}^{-1}\, (\widetilde{\ull{\bfg e}^{\jmath\,\theta}})^{-1}\, \mZmateq\, \ull{\bfg e}^{\jmath\,\theta^{+}}\,,\label{eq:exactsol0}\\
    \sone &= \widetilde{\ull{\bfg v}}^{-1}\, \left(\mZmateqp- \uletavecam \, \mZmateq\right)\, \ull{\bfg e}^{\jmath\,\theta^{+}}\,,\label{eq:exactsol1}\\
    \stwo &= \widetilde{\ull{\bfg v}}^{-1}\, \left(\mZmateqpp- \uletaavecam \, \mZmateq\right)\, \ull{\bfg e}^{\jmath\,\theta^{+}}\,,\label{eq:exactsol2}
\end{align}
where:
\begin{align}
    \mZmateq&=\left(\Iden - \mZmat\,  \ull{\bfg A}\right)^{-1}\, \mZmat\,,\label{eq:zeq}\\
    \mZmateqp&=\left(\Iden - \mZmat\,  \ull{\bfg B}'\right)^{-1}\, \mZmat\, \ull{\bfg A}'\, \mZmateq\,,\label{eq:zeqp}\\
    \mZmateqpp&=\left(\Iden - \mZmat\,  \ull{\bfg C}''\right)^{-1}\, \mZmat\,\left( \ull{\bfg A}''\, \mZmateq+\ull{\bfg B}''\, \mZmateqp\,\right)\,.\label{eq:zeqpp}
\end{align}

$\Iden$ is an identity matrix whose $k$-th diagonal element is $\M{1}{0}{0}{1}$, $\mZmat$ is the impedance matrix of the network, and $\ull{\bfg A}$, $\ull{\bfg A'}$, $\ull{\bfg A''}$, $\ull{\bfg B'}$, $\ull{\bfg B''}$, $\ull{\bfg C''}$ are matrices capturing the effect of the devices composing the network on strength.  We elaborate on the content of the latter matrices in Section \ref{subsec:devices}.

%Bold lowercase letters (e.g., $\boldsymbol{v}$) denote column vectors containing variables for every bus of the network.  Bold uppercase letters (e.g., $\boldsymbol{Z}$) denote square matrices containing variables that relate every pair of buses of the network.  The symbol $\sim$ over a quantity denotes the \textit{instantaneous arithmetic mean} as defined in the addendum of \cite{strength1}.  For instance, $\widetilde{x}=0.5\,(x^{+}+x^{-})$, where $x^+$ and $x^-$ denote the post- and pre-disturbance values of $x$, respectively.  
Quantities with $\sim$ in (\ref{eq:exactsol0})-(\ref{eq:exactsol2}) make the solutions dependent on the post-disturbance values, complicating the applicability of the set of metrics.  In this paper, we neglect this dependency by assuming the post-disturbance values are approximately equal to the pre-disturbance values, i.e., $x^{+}\approx x^{-}=x_0$.  \footnote{Hereinafter, the pre-disturbance values are denoted by subscript $0$ instead of superscript $-$, i.e, $x^{-}=x_0$.} With the assumption above, the solutions for the zero-, first- and second-order metrics simplify as follows:
\begin{align}
    \szero &= \ull{\bfg v_0}^{-1}\, \mZmateq\, \ull{\bfg e}^{\jmath\,\theta_{0}}\,,\label{eq:simplersol0}\\
    \sone &= \ull{\bfg v_0}^{-1}\, \mZmateqp\,  \ull{\bfg e}^{\jmath\,\theta_{0}}\,,\label{eq:simplersol1}\\
    \stwo &= \ull{\bfg v_0}^{-1}\, \mZmateqpp\, \ull{\bfg e}^{\jmath\,\theta_{0}}\,.\label{eq:simplersol2}
\end{align}

Note that $\ull{\bfg v_0}^{-1}=\bfg v_0^{-1}\circ\ull{\bfg e}^{-\jmath\,\theta_{0}}$ where $\circ$ denotes the Hadamard product (also known as element-wise product).  Focusing on the diagonal elements of $\szero$, $\sone$ and $\stwo$, it can be shown that the exponential terms $\ull{\bfg e}^{-\jmath\,\theta_{0}}$ and $\ull{\bfg e}^{\jmath\,\theta_{0}}$ cancel each other.  Consequently, (\ref{eq:simplersol0})-(\ref{eq:simplersol2}) can be further simplified as follows:
\begin{equation}\label{eq:finalsol}
\boxed{
    \szero = \bfg v_0^{-1}\, \mZmateq \,,\quad
    \sone = \bfg v_0^{-1}\, \mZmateqp \,,\quad
    \stwo = \bfg v_0^{-1}\, \mZmateqpp\,.
}
\end{equation}

The solution given by (\ref{eq:finalsol}) is the final simplified solution for the whole set of strength metrics.  It depends on the impedance matrix $\mZmat$, devices' contribution to strength $\ull{\bfg A}$, $\ull{\bfg A'}$, $\ull{\bfg A''}$, $\ull{\bfg B'}$, $\ull{\bfg B''}$, $\ull{\bfg C''}$, and the pre-fault voltage $\bfg{v_0}$.  In case the AC power flow solution is not available (needed for ${\bfg v_0}$), the DC power flow solution can be taken, i.e., all voltage magnitudes equal to 1.  The effect of this simplification is a small scaling in the values of the metrics, which leads to an underestimation of the level of strength for the buses with over-voltages and an overestimation for those with under-voltages.  Alternatively, limit case factors can be used, e.g., 0.9 leads to the weakest values whereas 1.1 leads to the strongest values.  These assumptions are essentially equivalent to well-established methods for calculating maximum/minimum short-circuit currents, e.g.  IEC 60909 \cite{iec60909}.
\subsection{The contribution of devices to strength}\label{subsec:devices}
As introduced in \cite{strength1} and without lack of generality, it is assumed that devices are modeled in a current-injection form, where a sudden change in the voltage---and its time derivatives---will produce a sudden change in their current injection---and its time derivatives as well---according to the following equations: 
\begin{align}
    \Delta\ul{\imath}_{\rm dev}&=\ull{a}\,\Delta\ul{v}\,,\\
    \Delta\dot{\ul{\imath}}_{\rm dev}&=\ull{a}'\, \Delta\ul{v}+\ull{b}'\, \Delta\dot{\ul{v}}\,,\\
    \Delta\ddot{\ul{\imath}}_{\rm dev}&=\ull{a}''\, \Delta\ul{v}+\ull{b}''\, \Delta\dot{\ul{v}}+\ull{c}''\, \Delta\ddot{\ul{v}}\,.
\end{align}

The parameters $\ull{a}$, $\ull{a}'$, $\ull{a}''$, $\ull{b}'$, $\ull{b}''$ and $\ull{c}''$ are the so-called strength components of devices, which depend on the variables and parameters of the device.  They capture their effect on strength and are used to populate the diagonal elements of $\ull{\bfg A}$, $\ull{\bfg A'}$, $\ull{\bfg A''}$, $\ull{\bfg B'}$, $\ull{\bfg B''}$, $\ull{\bfg C''}$ at the positions corresponding to their terminal bus.  In case there is multiple devices connected at the same bus, their strength components add up.

While the strength components of devices are very convenient for calculating the whole network strength metrics using (\ref{eq:finalsol}), it is hard to interpret the role each one of them plays on strength.  To address this issue, we present an alternative form obtained from calculating the strength metrics at the terminal bus of a device in open-circuit conditions.  In other words, given an isolated device with certain strength components, the strength metrics at its terminal bus are as follows:

\begin{equation}\label{eq:altform}
\begin{aligned}
    \ull{S}&=-\ull{a}^{-1}\\
    \ull{S}'&=\ull{b}'^{-1}\cdot\ull{a}'\cdot \ull{a}^{-1}\\
    \ull{S}''&=\ull{c}''^{-1}\cdot \left(\ull{a}''-\ull{b}''\cdot\ull{b}'^{-1}\cdot \ull{a}'\right)\cdot\ull{a}^{-1}
\end{aligned}
\end{equation}

From (\ref{eq:altform}) it can be noted that the zero-order strength level depends solely on the zero-order abc component $\ull{a}$.  In turn, the higher order strength levels depend on the strength components of the same order, but also on the components of the lower levels.  For example, $\ull{S}'$ depends on $\ull{a}$, and
$\ull{S}''$ depends on $\ull{a}'$, $\ull{b}'$, and $\ull{a}$.

Analytical expressions for the strength components of specific devices can be derived by following a systematic procedure.  Starting from the set of DAEs of the device model, the first step is to find expressions for the device current injection as a function of model inputs, states, and the terminal voltage.  By applying the $\Delta$ operator to the equations found, the strength components of the device can be identified.  Detailed step-by-step examples for a variety of device models are provided in \cite{strength1}.  In the appendix of this paper, Tables \ref{tab:syn}, \ref{tab:zload}, and \ref{tab:gfl} summarize the resulting strength components for a group of relevant devices.  Compared to those in \cite{strength1}, the expressions herein presented are simpler and therefore more practical.  Simplifications are based on the same assumption described in Section \ref{subsec:simple}.  Note that, while the `diagonal' strength components, i.e., $\ull{a}$, $\ull{b'}$ and $\ull{c}''$, are never null, the off-diagonal components can be null in some cases.  This is important since, as it can be noted from (\ref{eq:altform}), if $\ull{a}'$ is null then $\ull{S}'$ is also null, meaning an infinitely strong first-order strength level.  Similarly, if $\ull{a}'$ and $\ull{a}''$ are null, the same happens to the second-order strength level $\ull{S}''$.  Consider a conventional system composed entirely by synchronous machines and loads.  Since $\ull{a}'$ is null for both devices, the first-order strength metrics is null for the whole network.  Consequently, the complex frequency cannot jump, i.e., it becomes a continuous variable of the system.  In turn, given the $\ull{a}''$ component of synchronous machines, the second-order strength level of the system is not null, i.e., the rate of change of the complex frequency changes after a sudden current change.  These observations are consistent with the well-known characteristics of the frequency dynamics of conventional systems.  In case a device introduces a non-null $\ull{a}'$ component, it will challenge the continuous nature of the first-order complex frequency.

\section{Dynamic analysis}\label{sec:dynamic}
In this section we aim to study the relationship between strength and power system dynamic performance.  The analysis provides valuable insights for defining acceptable strength levels based on our framework.
\subsection{Strength source model}\label{subsec:source}

To study the impact of strength on the dynamic performance of power-system devices, it is necessary to vary the strength levels to which devices are exposed in a controlled and independent manner.  To address this challenge, we introduce a novel strength source model that emulates an external equivalent network with prescribed strength levels.  The model is formulated as a second-order linear dynamic system capable of reproducing arbitrary zero-, first-, and second-order strength metrics at its terminals.  Thus, 
the twelve strength metrics at the connection point can be specified as parameters $\ull{S}$, $\ull{S}'$, and $\ull{S}''$:
\begin{align}
    \ull{S}&=\M{S_{v\imath_{\rm p}}}{S_{v\imath_{\rm q}}}{S_{\theta\imath_{\rm p}}}{S_{\theta\imath_{\rm p}}},\, \ull{S}'=\M{S_{\rho\imath_{\rm p}}}{S_{\rho\imath_{\rm q}}}{S_{\omega\imath_{\rm p}}}{S_{\omega\imath_{\rm p}}},\,\ull{S}''=\M{S_{\sigma\imath_{\rm p}}}{S_{\sigma\imath_{\rm q}}}{S_{\gamma\imath_{\rm p}}}{S_{\gamma\imath_{\rm p}}}
\end{align}

In the absence of any other device in the network, the actual strength level of the connection bus will match exactly those of the source.

The set of differential equations is as follows:
\begin{align}
\dot{m}_{\rm d}&=K_{\rm 00}v_{\rm d}+K_{\rm 01}v_{\rm q}+m'_{\rm d}\,,\\
\dot{m}_{\rm q}&=K_{\rm 10}v_{\rm d}+K_{\rm 11}v_{\rm q}+m'_{\rm q}\,,\\
\dot{m}'_{\rm d}&=L_{\rm 00}v_{\rm d}+L_{\rm 01}v_{\rm q}+m''_{\rm d0}\,,\\
\dot{m}'_{\rm q}&=L_{\rm 10}v_{\rm d}+L_{\rm 11}v_{\rm q}+m''_{\rm q0}\,,
\end{align}
where:
{
\setlength{\arraycolsep}{1.5pt}
\begin{align}
\begin{bmatrix}
    K_{\rm00}&K_{\rm01}\\
    K_{\rm10}&K_{\rm11}
\end{bmatrix}
&=
\begin{bmatrix}
    S_{\rho\imath_{\rm p}}&S_{\rho\imath_{\rm q}}\\
    S_{\omega\imath_{\rm p}}&S_{\omega\imath_{\rm q}}
\end{bmatrix}
\cdot
\begin{bmatrix}
    S_{v\imath_{\rm p}}&S_{v\imath_{\rm q}}\\
    S_{\theta\imath_{\rm p}}&S_{\theta\imath_{\rm q}}
\end{bmatrix}^{-1}\,,\\
\begin{bmatrix}
    L_{\rm00}&L_{\rm01}\\
    L_{\rm10}&L_{\rm11}
\end{bmatrix}
&=
\begin{bmatrix}
    S_{\sigma\imath_{\rm p}}&S_{\sigma\imath_{\rm q}}\\
    S_{\gamma\imath_{\rm p}}&S_{\gamma\imath_{\rm q}}
\end{bmatrix}
\cdot
\begin{bmatrix}
    S_{v\imath_{\rm p}}&S_{v\imath_{\rm q}}\\
    S_{\theta\imath_{\rm p}}&S_{\theta\imath_{\rm q}}
\end{bmatrix}^{-1}
+
\begin{bmatrix}
    K_{\rm00}&K_{\rm01}\\
    K_{\rm10}&K_{\rm11}
\end{bmatrix}^{2}\,.
\nonumber
\end{align}
}
The strength source model allows testing the dynamic behavior of different devices under arbitrary strength levels at their point of connection.

\subsection{Bus-level versus device-level strength metrics}\label{subsec:devicelevel}
By construction, the formulation (\ref{eq:f0})-(\ref{eq:f2}) defines a set of bus-level metrics, which represent the strength level of a bus in \textit{absolute} terms, i.e., their value is the actual sensitivity of the three-phase AC voltage of a bus to the current injection at the bus.  In order to judge if the numerical value of a metric should be considered `strong' or `weak', it has to be compared against the size of the expected current change.  From the point of view of a device, it can be compared with the rated current of surrounding equipment.  A similar interpretation applies to the short-circuit level and the SCR.  The short-circuit level is an absolute metric reflecting the sensitivity of the voltage magnitude to reactive power variations.  However, the same value may indicate a `strong' or `weak' condition depending on the rating of the connected equipment.  In contrast, the SCR provides a relative measure by relating the short-circuit level to the device size, thereby enabling the identification of operating ranges typically associated with `strong' or `weak' conditions.

The discussion above motivates the definition of ratio-like quantities for representing device-level strength based on our formulation.  Consider a device with base current $\imath_{\rm b}$.  The ratio-like strength metrics for the device are defined as:
{\renewcommand{\arraystretch}{1.2}
\begin{align}
    \ull{S\scriptstyle \rm R}&=\left[\begin{array}{cc}S\scriptstyle{\rm R}_{\scriptstyle v\imath_{\rm p}}&S\scriptstyle{\rm R}_{\scriptstyle v\imath_{\rm q}}\\
        S\scriptstyle{\rm R}_{\scriptstyle\theta\imath_{\rm p}}&S\scriptstyle{\rm R}_{\scriptstyle\theta\imath_{\rm q}}\end{array}\right]:=\frac{1}{\imath_{\rm b}}\left[\begin{array}{cc}S_{v\imath_{\rm p}}^{-1}&S_{v\imath_{\rm q}}^{-1}\\
        S_{\theta\imath_{\rm p}}^{-1}&S_{\theta\imath_{\rm q}}^{-1}\end{array}\right]\,,\\
    \ull{S\scriptstyle \rm R}'&=\left[\begin{array}{cc}S\scriptstyle{\rm R}_{\scriptstyle\rho\imath_{\rm p}}&S\scriptstyle{\rm R}_{\scriptstyle\rho\imath_{\rm q}}\\
        S\scriptstyle{\rm R}_{\scriptstyle\omega\imath_{\rm p}}&S\scriptstyle{\rm R}_{\scriptstyle\omega\imath_{\rm q}}\end{array}\right]:=\frac{1}{\imath_{\rm b}}\left[\begin{array}{cc}S_{\rho\imath_{\rm p}}^{-1}&S_{\rho\imath_{\rm q}}^{-1}\\
        S_{\omega\imath_{\rm p}}^{-1}&S_{\omega\imath_{\rm q}}^{-1}\end{array}\right]\,,\\
    \ull{S\scriptstyle \rm R}''&=\left[\begin{array}{cc}S\scriptstyle{\rm R}_{\scriptstyle\sigma\imath_{\rm p}}&S\scriptstyle{\rm R}_{\scriptstyle\sigma\imath_{\rm q}}\\
        S\scriptstyle{\rm R}_{\scriptstyle\gamma\imath_{\rm p}}&S\scriptstyle{\rm R}_{\scriptstyle\gamma\imath_{\rm q}}\end{array}\right]:=\frac{1}{\imath_{\rm b}}\left[\begin{array}{cc}S_{\sigma\imath_{\rm p}}^{-1}&S_{\sigma\imath_{\rm q}}^{-1}\\
        S_{\gamma\imath_{\rm p}}^{-1}&S_{\gamma\imath_{\rm q}}^{-1}\end{array}\right]\,,
\end{align}
where} (${S}_{v\imath_{\rm p}}$, ${S}_{v\imath_{\rm q}}$, ${S}_{\theta\imath_{\rm p}}$ and ${S}_{\theta\imath_{\rm q}}$), (${S}_{\rho\imath_{\rm p}}$, ${S}_{\rho\imath_{\rm q}}$, ${S}_{\omega\imath_{\rm p}}$ and ${S}_{\omega\imath_{\rm q}}$), and (${S}_{\sigma\imath_{\rm p}}$, ${S}_{\sigma\imath_{\rm q}}$, ${S}_{\gamma\imath_{\rm p}}$ and ${S}_{\gamma\imath_{\rm q}}$) are the zero-, first- and second-order strength levels of the terminal bus of the device, respectively.

Normalization of the bus-level metrics is achieved by multiplying them by the rated current of the device.  Even though that would suffice for defining the relative metrics, we also take the inverse so that a higher value indicates a stronger condition.  By doing so, our ratio-like metrics follow the same convention as for SCR.  The relative metrics defined above enable the search for correlations between strength and stability, and the identification of typical ranges for strong or weak conditions.

\subsection{Single device vs strength source}\label{subsec:sdas}
\begin{figure}[hbtp]

    \centering
    \includegraphics[width=0.6\linewidth]{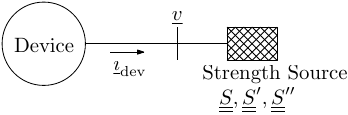}

    \caption{Single device against the strength source.}
    \label{fig:sdas}
\end{figure}

In this section, we set up a benchmark system composed of a single device against the strength source model introduced in \ref{subsec:source}, as illustrated in Fig.  \ref{fig:sdas}.  The purpose is to study how the dynamic performance of a device changes when subjected to different strength levels.  

\begin{figure}[t]
    \centering
    % Top figure
    \includegraphics[width=0.9\linewidth]{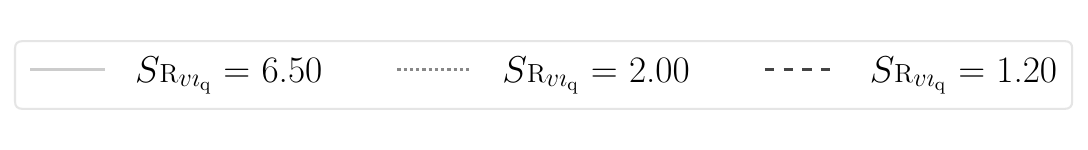}
    % Bottom left
    \begin{minipage}{0.49\linewidth}
        \centering
        \includegraphics[width=\linewidth]{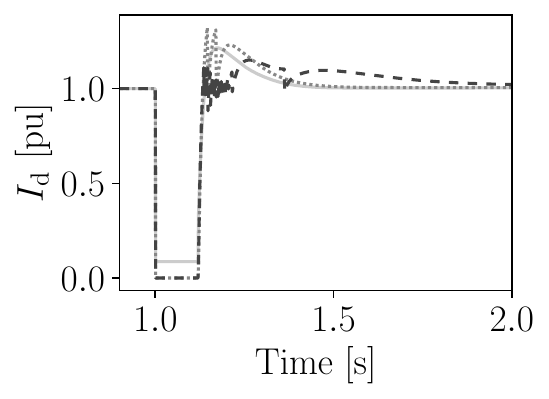}
    \end{minipage}
    \hfill
    % Bottom right
    \begin{minipage}{0.49\linewidth}
        \centering
        \includegraphics[width=\linewidth]{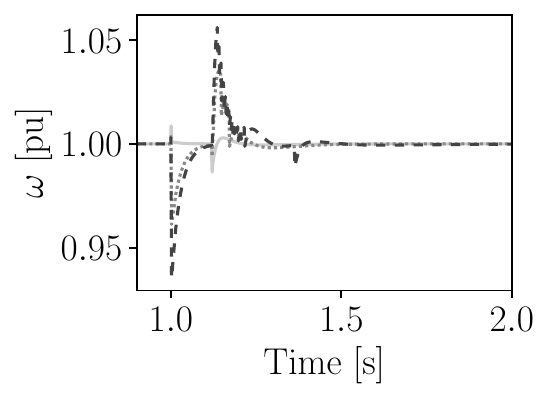}
    \end{minipage}
    \caption{Direct axis current of the converter (left panel) and Converter's PLL frequency (right panel) for varying strength ratios $S\scriptstyle{\rm R}_{\scriptstyle v\imath_{\rm q}}$.}
    \label{fig:Id_s01}
\end{figure}

\begin{figure}[t]
    \centering
    % Top figure
    \includegraphics[width=0.9\linewidth]{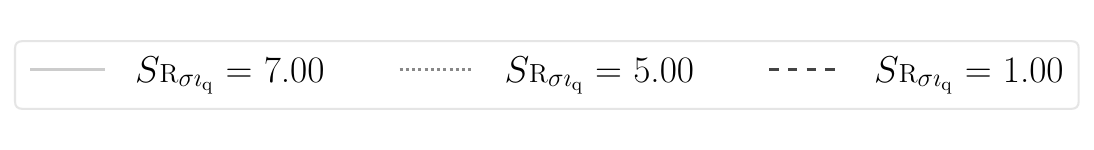}
    % Bottom left
    \begin{minipage}{0.49\linewidth}
        \centering
        \includegraphics[width=\linewidth]{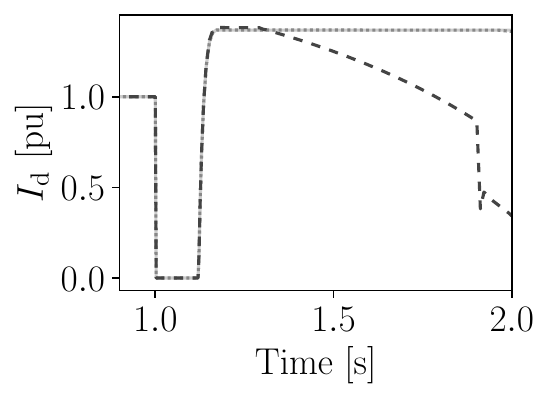}
    \end{minipage}
    \hfill
    % Bottom right
    \begin{minipage}{0.49\linewidth}
        \centering
        \includegraphics[width=\linewidth]{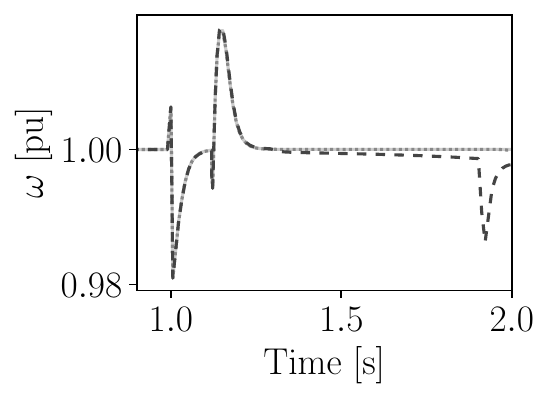}
    \end{minipage}
    \caption{Direct axis current of the converter (left panel) and Converter's PLL frequency (right panel) for varying strength ratios $S\scriptstyle{\rm R}_{\scriptstyle \sigma\imath_{\rm q}}$.}
    \label{fig:Id_spp01}
\end{figure}

Consider a GFL converter modeled according to the generic WECC/EPRI renewable energy models REGCA1, REECA1, and REPCA1 \cite{EPRI_WECC_GenericModels}.  We sweep a range of values for each strength parameter using the strength source, keeping the rest of them constant.  For each condition, a time-domain simulation of a three-phase short-circuit at the device's terminal bus is performed.  Fig.  \ref{fig:Id_s01} depicts the results for the sweep in the zero-order strength metric $S\scriptstyle{\rm R}_{\scriptstyle v\imath_{\rm q}}$.  As expected, converter's dynamic performance greatly degrades for low strength ratios below 2.0.  Similarly, Fig.  \ref{fig:Id_spp01} depicts the results for the sweep in the second-order strength metric $S\scriptstyle{\rm R}_{\scriptstyle \sigma\imath_{\rm q}}$.  In this case, the dynamic performance is degraded in a different way, manifested in a voltage collapse for a strength ratio below 5.0.  

%\begin{figure}[hbtp]
%    \centering
%    \includegraphics[width=0.84\linewidth]{Images/Id_s01.pdf}
%    \caption{Direct axis current of the converter for varying strength ratios %$S\scriptstyle{\rm R}_{\scriptstyle v\imath_{\rm q}}$.}
%    \label{fig:Id_s01}
%\end{figure}
%\begin{figure}[hbtp]
%    \centering
%    \includegraphics[width=0.84\linewidth]{Images/omega_s01.pdf}
%    \caption{Converter's PLL frequency for varying strength ratios $S\scriptstyle{\rm %R}_{\scriptstyle v\imath_{\rm q}}$.}
%    \label{fig:omega_s01}
%\end{figure}

%\begin{figure}[hbtp]
%    \centering
%    \includegraphics[width=0.84\linewidth]{Images/Id_spp01.pdf}
%    \caption{Direct axis current of the converter for varying strength ratios %$S\scriptstyle{\rm R}_{\scriptstyle \sigma\imath_{\rm q}}$.}
%    \label{fig:Id_spp01}
%\end{figure}
%\begin{figure}[hbtp]
%    \centering
%    \includegraphics[width=0.84\linewidth]{Images/omega_spp01.pdf}
%    \caption{Converter's PLL frequency for varying strength ratios $S\scriptstyle{\rm %R}_{\scriptstyle \sigma\imath_{\rm q}}$.}
%    \label{fig:omega_spp01}
%\end{figure}

The same procedure can be done for the complete set of strength metrics in a range of events and operating conditions, resulting in the minimum strength limits the device needs for proper operation.  The analysis provided in this section illustrates how strength limits can be found based on our formulation and strength source.

\section{Study Case}\label{sec:studycase}
In this section we implement our framework in a real study case, the All-Island Irish power system.
\subsection{System model}
Ireland's power system is represented using a positive-sequence dynamic model adapted from the All-Island Ten Year Transmission Forecast Statement 2022 study files, published by the Irish TSOs at \cite{EirGrid2022TYTFS}.  

The model is composed of 2080 buses and 1176 lines.  We consider a summer peak scenario with 5{,}740 MW demand and 32\% System Non-Synchronous Penetration (SNSP).
% Fig.  \ref{fig:txmap} depicts a map of the transmission system.
Dynamic models used are based on PSS/E library models.  Strength-relevant devices are synchronous generators, loads, and grid-following converters.  
All modeling and simulations are conducted in the Python-based software tool Dome \cite{dome}.  

The results presented in this section are intended for methodological illustration only and do not represent an assessment of the actual strength of the Irish transmission system.  The obtained strength levels depend on the adopted modeling assumptions and operating scenario.

%\begin{figure}[hbtp]

%    \centering
%    \includegraphics[width=0.9\linewidth, trim={1cm 2.0cm 1cm 2.5cm},clip]{Images/ireland_map.pdf}

%    \caption{All-Island Irish transmission system map.}
%    \label{fig:txmap}
%\end{figure}

\subsection{Bus-level strength metrics results}
\begin{figure}[hbtp]
\centering
\includegraphics[width=0.9\linewidth]{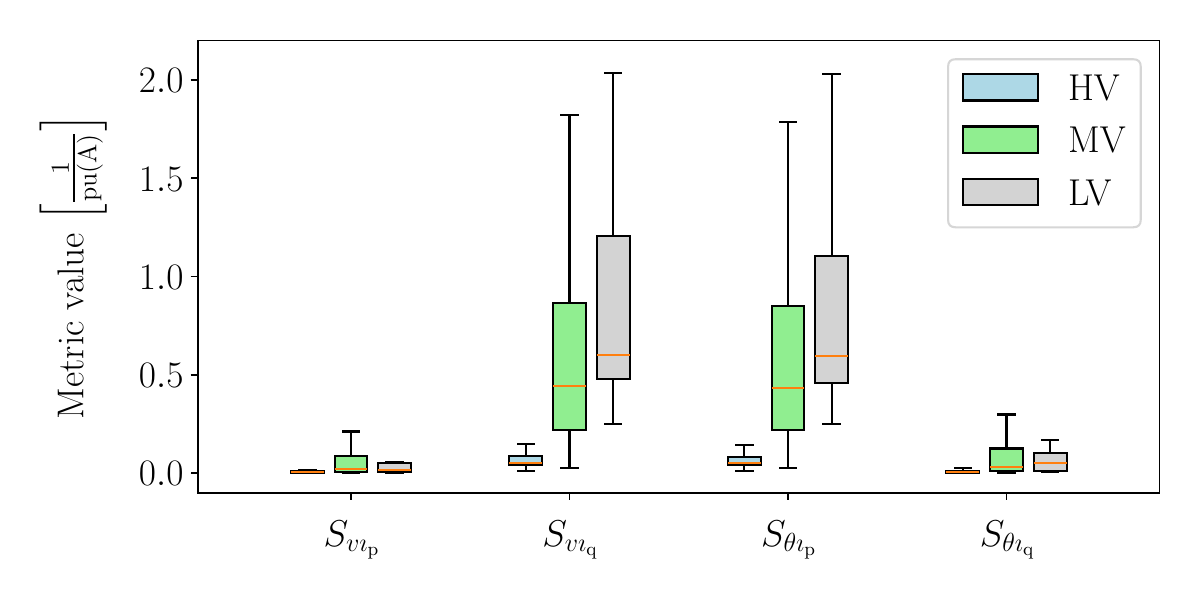}
\caption{Zero-order strength metric results.}
\label{fig:box0}
\end{figure}

We implement the framework described in Section \ref{sec:framework} and calculate the set of twelve strength metrics for all buses of the system.  Fig.  \ref{fig:box0} shows the results for the zero-order metrics in box and whisker plots\footnote{The box extends from the first and third quartiles, with an orange line at the median.  The whiskers extend from the box to the farthest data point lying within 1.5x the inter-quartile range from the box.}.  Results are grouped for high-voltage (HV)  $V_{\rm b}\geq 110$ kV, medium-voltage (MV) $10\text{ kV}<V_{\rm b}\leq 110$ kV and low-voltage buses (LV) $V_{\rm b}< 10$ kV.  As expected, the sensitivity of the voltage magnitude is greater to the reactive current than to the active current, i.e., $S_{v\imath_{\rm q}}>S_{v\imath_{\rm p}}$.  Similarly, the sensitivity of the voltage angle to the active current is greater than the sensitivity to the reactive current, i.e., $S_{\theta\imath_{\rm p}}>S_{\theta\imath_{\rm q}}$.  In addition, buses of a higher voltage level are generally stronger than lower voltage level buses.  Considering the full set of results, zero-order strength metrics typically range from 0.01 $\rm pu^{-1}$ to 2.0 $\rm pu^{-1}$, where the base current is such that the power base is 100 MVA.  At the transmission level, the results for HV buses range from 0.01 $\rm pu^{-1}$ to 0.3 $\rm pu^{-1}$.  Leveraging the clear physical domain of our metrics, this range indicates that for a sudden current change equivalent to a 100 MVA power change, the voltage at HV buses will change from 0.01 $\rm pu$ for the strongest buses to 0.3 $\rm pu$ for the weakest buses.

All first-order metrics are null for the entire network as there is no source of $a'$ component from any of the devices in the system.  As explained in Section \ref{subsec:devices}, in this situation the complex frequency becomes a continuous variable.

\begin{figure}[hbtp]
\centering
\includegraphics[width=0.9\linewidth]{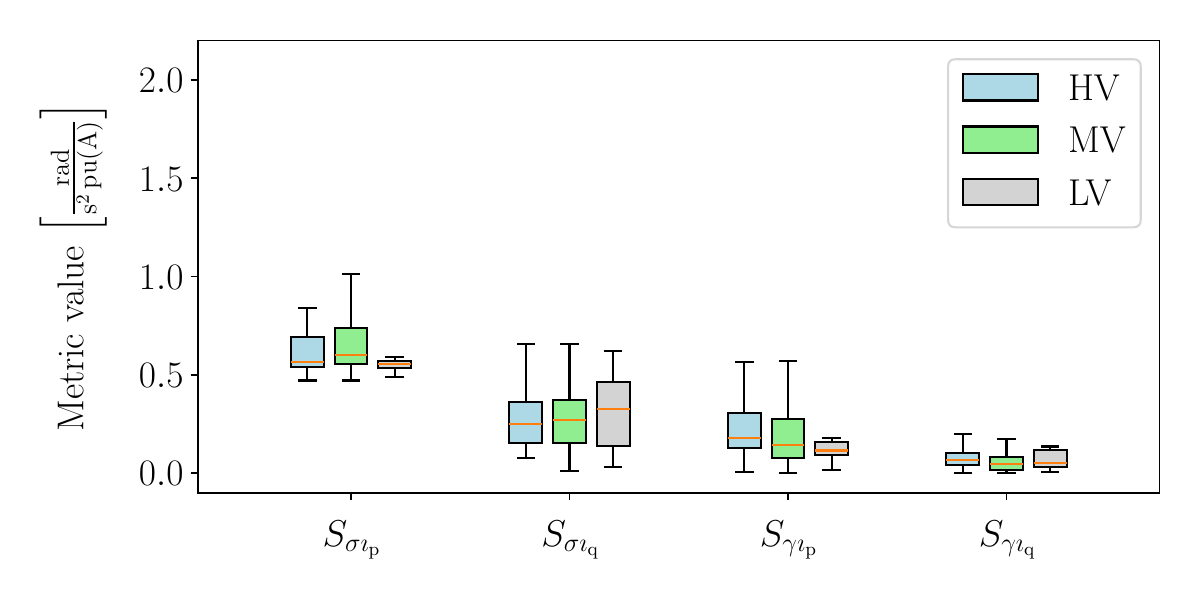}
\caption{Second-order strength metric results.}
\label{fig:box2}
\end{figure}

Fig.  \ref{fig:box2} shows the results for the second-order metrics.  In this case, the four metrics span across a narrower range, being $S_{\sigma\imath_{\rm p}}$ and $S_{\gamma\imath_{\rm q}}$ the ones showing the highest and smallest values, respectively.  Remarkably---and unlike the zero-order metrics---, results are fairly homogeneous across different voltage levels, i.e., the rate of change of the complex frequency is not weaker for LV buses.  Focusing on $S_{\gamma\imath_{\rm p}}$, a sudden current change equivalent to a 100 MW power change will cause the RoCoF to reach a value around 0.1 $\rm rad/s^2$ (0.02 $\rm Hz/s$) for the strongest buses to 1.2 $\rm rad/s^2$ (0.2 $\rm Hz/s$) for the weakest buses.

In order to analyze the spatial distribution of the metrics, the results for HV buses (above 110 kV) are shown in geographic diagrams using a grayscale, where a darker color indicates a weaker value.

\begin{figure}[hbtp]
    \centering
    \includegraphics[width=\linewidth, trim={0 1cm 0 0}, clip]{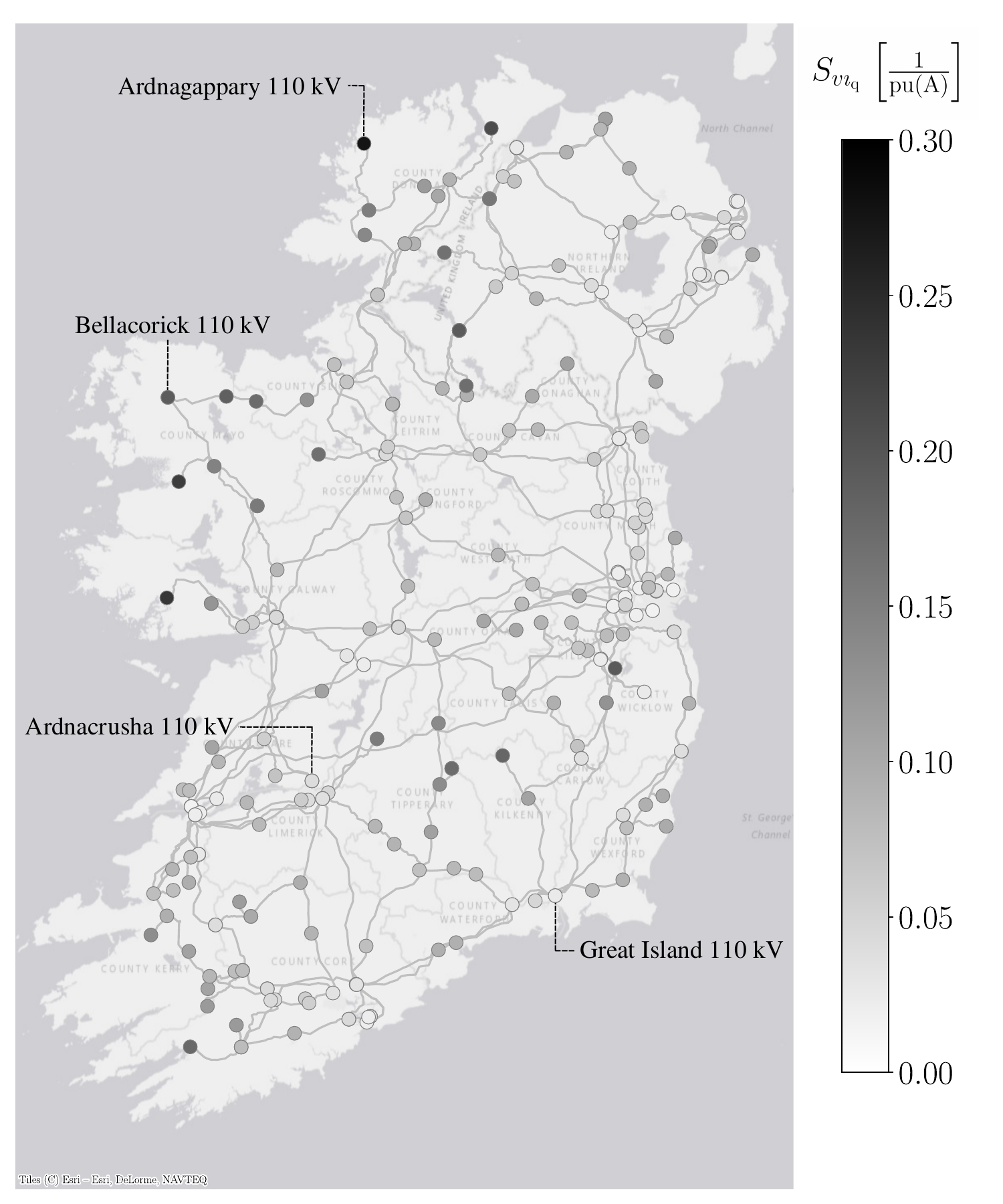}
    \caption{All-Island Irish power system results for strength metric $S_{v\imath_{\rm q}}$.}
    \label{fig:MAP_S_viq}
\end{figure}

Fig.  \ref{fig:MAP_S_viq} depicts the results for the zero-order metric $S_{v\imath_{\rm q}}$.  As expected, buses in low-meshed areas tend to be relatively weaker.  This metric greatly correlates to the short-circuit level of the bus.  Results for the zero-order metric $S_{\theta\imath_{\rm p}}$ are approximately equal to those shown in Fig.  \ref{fig:MAP_S_viq}.  The slight difference is due to the relatively small asymmetry in the dq-axis transient reactances of synchronous machines.  Finally, the other two zero-order metrics $S_{v\imath_{\rm p}}$ and $S_{\theta\imath_{q}}$ are one order of magnitude stronger compared to $S_{v\imath_{\rm q}}$ and $S_{\theta\imath_{\rm p}}$.  This is the typical case for HV transmission systems where $r\ll x$ across the system.

\begin{figure}[hbtp]
    \centering
    \includegraphics[width=\linewidth, trim={0 1cm 0 0}, clip]{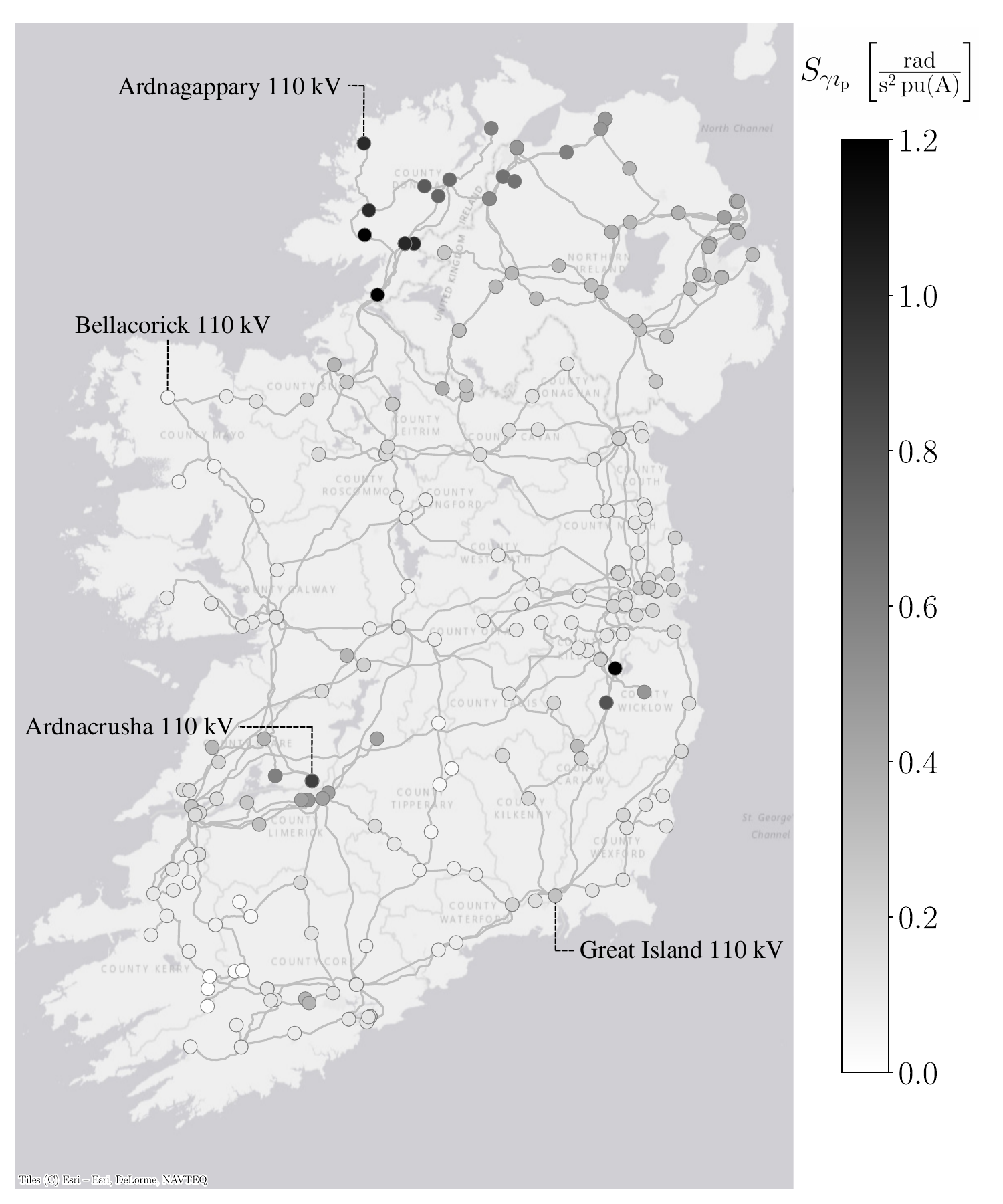}
    \caption{All-Island Irish power system results for strength metric $S_{\gamma\imath_{\rm p}}$.}
    \label{fig:MAP_S_gammaip}
\end{figure}

\begin{figure}[hbtp]
    \centering
    \includegraphics[width=\linewidth, trim={0 1cm 0 0}, clip]{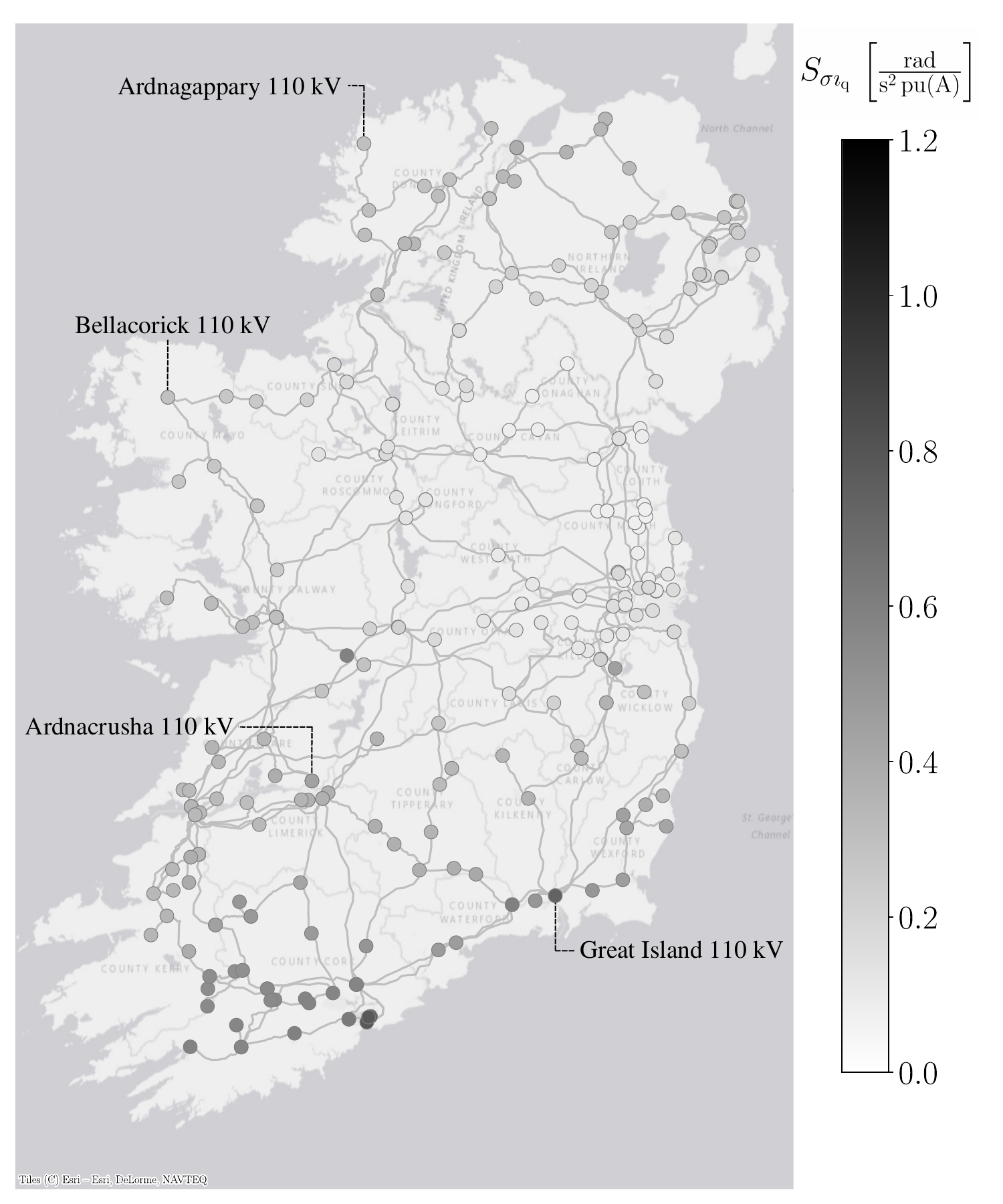}
    \caption{All-Island Irish power system results for strength metric $S_{\sigma\imath_{\rm q}}$.}
    \label{fig:MAP_S_sigmaiq}
\end{figure}

Fig.  \ref{fig:MAP_S_gammaip} depicts the results for the second-order metric $S_{\gamma\imath_{\rm p}}$.  Buses in the vicinity of synchronous generators show the weakest values.  Their RoCoF dynamic is greatly dominated by the relatively low local inertia, while buses furthest from generators leverage the inertial contribution from more machines.  Fig.  \ref{fig:MAP_S_sigmaiq} shows the results for the second-order metric $S_{\sigma\imath_{\rm q}}$.  The second-order rate of change of the voltage, i.e., $\dot{\rho}$, turns out to be weaker in the South-West of the Republic of Ireland, and moderately weak in the North-West of Northern Ireland.  In both regions there is a high presence of wind farms that are operating in a constant reactive power control mode.  Finally, the other two second-order indicators $S_{\gamma\imath_{\rm q}}$ and $S_{\sigma\imath_{\rm p}}$ show a relatively homogeneous distribution across the grid.

Simulation results show that substations might be relatively strong for certain metrics and weak for other metrics at the same time.  
For example, there is buses with strong zero-order metrics and weak second-order metrics, and vice-versa.  Similarly, we find strong buses in terms of the second order metric $S_{\gamma\imath_{\rm p}}$ but weak in terms of $S_{\sigma\imath_{\rm p}}$, among other combinations.  The voltage deviation after a perturbation is characterized according to the strength level the bus shows for the different metrics.  Some special cases are summarized in Table \ref{tab:spec}.  

\begin{table}[hbtp]
\renewcommand{\arraystretch}{1.2}
\caption{Special cases with different strength levels per metric.}
\centering
\begin{tabular}{c|cc|cc}
\hline
\multirow{2}{*}{Substation} & \multicolumn{2}{c|}{Zero-order strength} & \multicolumn{2}{c}{Second-order strength} \\
        & $S_{v\imath_{\rm q}}$  & $S_{\theta\imath_{\rm p}}$     & $S_{\gamma\imath_{\rm p}}$      & $S_{\sigma\imath_{\rm q}}$    \\ \hline
Bellacorick 110 kV   & \multicolumn{2}{c|}{Weak}      & Strong        & Moderate               \\
Ardnacrusha 110 kV  & \multicolumn{2}{c|}{Strong}    & Weak        & Moderate               \\
Great Island 110 kV   & \multicolumn{2}{c|}{Strong}      & Moderate        & Weak               \\ 
Ardnagappary 110 kV  & \multicolumn{2}{c|}{Weak}      & Weak        & Moderate               \\ \hline
\end{tabular}
\label{tab:spec}
\end{table}

\begin{figure}[t]
    \centering
    % Top figure
    \includegraphics[width=\linewidth]{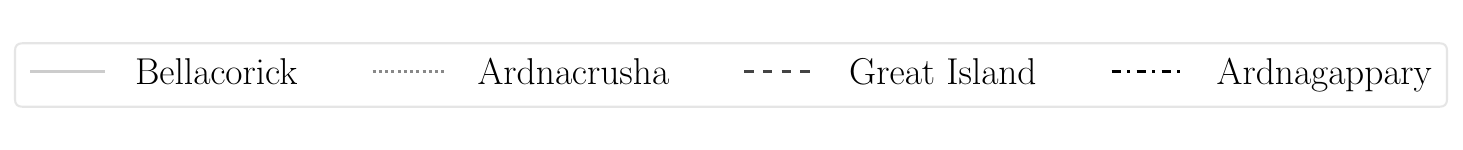}
    % Bottom left
    \begin{minipage}{0.49\linewidth}
        \centering
        \includegraphics[width=\linewidth]{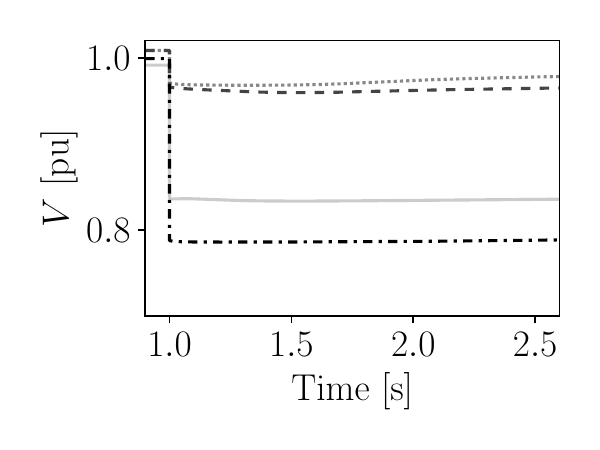}
    \end{minipage}
    \hfill
    % Bottom right
    \begin{minipage}{0.49\linewidth}
        \centering
        \includegraphics[width=\linewidth]{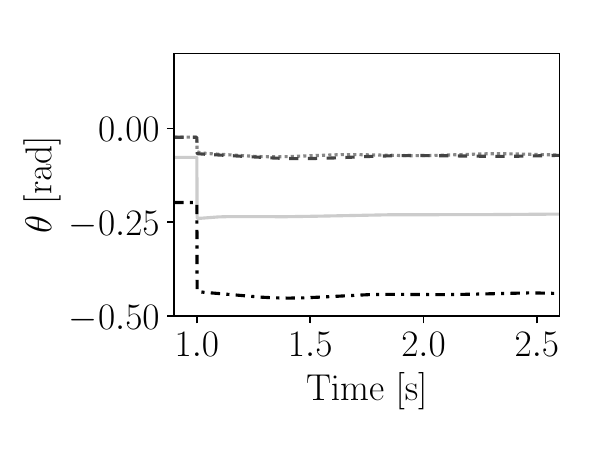}
    \end{minipage}

    \caption{Trajectories of the voltage magnitude (left panel) and voltage phase (right panel) for a sudden change of power at different substations.}
    \label{fig:voltage}
\end{figure}

\begin{figure}[t]
    \centering
    % Top figure
    \includegraphics[width=\linewidth]{Images/legend.pdf}
    % Bottom left
    \begin{minipage}{0.49\linewidth}
        \centering
        \includegraphics[width=\linewidth]{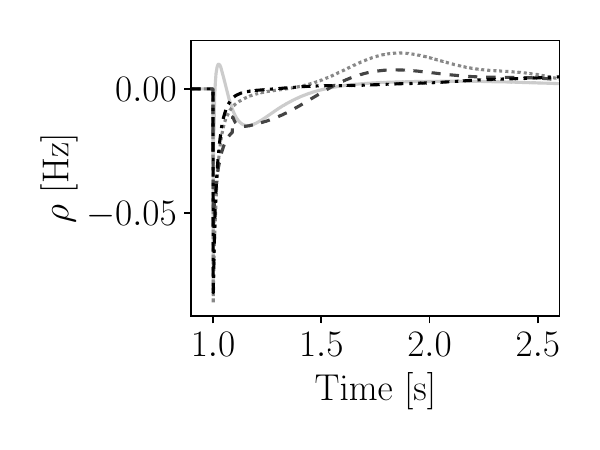}
    \end{minipage}
    \hfill
    % Bottom right
    \begin{minipage}{0.49\linewidth}
        \centering
        \includegraphics[width=\linewidth]{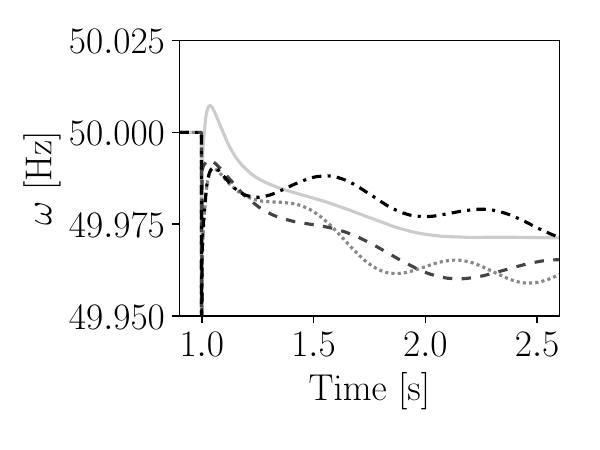}
    \end{minipage}

    \caption{Trajectories of the rate of change of voltage (left panel) and frequency (right panel) for a sudden change of power at different substations.}
    \label{fig:omega}
\end{figure}

For each substation in Table \ref{tab:spec}, time-domain simulations are performed for a sudden change of active and reactive power equivalent to 100 MVA.  The trajectories of the voltage magnitude and phase are depicted in Fig.  \ref{fig:voltage}.  Results show well that the sudden drop in the voltage magnitude and phase is significantly higher for the zero-order weak substations, i.e., Bellacorick 110 kV and Ardnagappary 110 kV.  The right panel of Fig.  \ref{fig:omega} depicts the trajectory of the imaginary part of the complex frequency, i.e., $\omega$ for the same substations.  Despite Bellacorick 110 kV has a weak zero-order response, its second-order strength is relatively higher.  Consequently, its frequency experiences a lower rate of change and less oscillations.  In turn, the other three substations exhibit a worse frequency trajectory due to their moderate to weak metric $S_{\gamma\imath_{\rm p}}$.  Similar conclusions can be drawn from the left panel of Fig.  \ref{fig:omega}, which depicts the trajectory of the real part of the complex frequency, i.e., $\rho$.  In this case, Great Island 110 kV experiences the highest rate of change of $\rho$, which aligns with its relatively weak metric $S_{\sigma\imath_{\rm q}}$.  By examining the resulting set of metrics of a bus, one can predict basic trajectory patterns its three-phase AC voltage will follow after a disturbance.

\subsection{Device-level strength metrics results}
\begin{figure}[t]
    \centering
    % Top figure
    \includegraphics[width=0.8\linewidth]{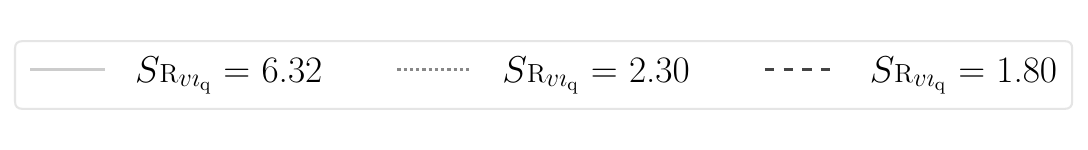}
    % Bottom left
    \begin{minipage}{0.49\linewidth}
        \centering
        \includegraphics[width=\linewidth]{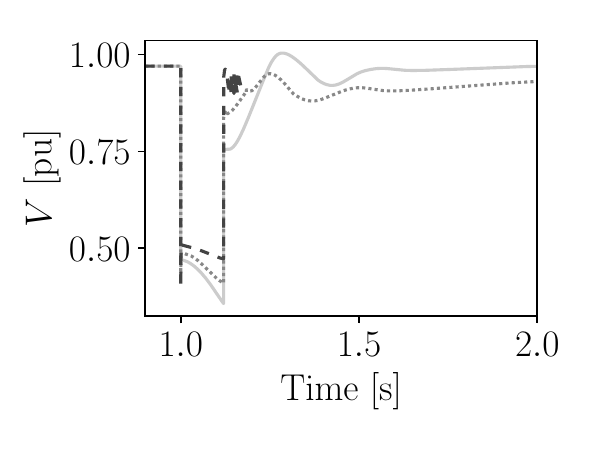}
    \end{minipage}
    \hfill
    % Bottom right
    \begin{minipage}{0.49\linewidth}
        \centering
        \includegraphics[width=\linewidth]{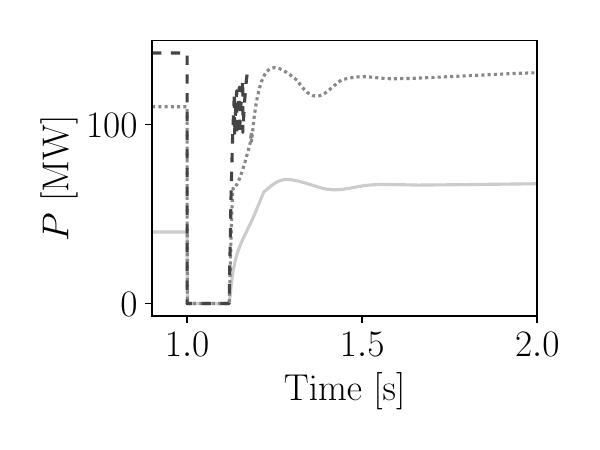}
    \end{minipage}

    \caption{Trajectories of the terminal voltage (left panel) and active power (right panel) from WF Knockacummer for different strength levels.}
    \label{fig:voltage_device}
\end{figure}

The bus-level strength metrics obtained above are combined with the current ratings of the devices in the network to calculate the ratio-like device-level metrics, as defined in Section \ref{subsec:devicelevel}.  Table \ref{tab:weakdev} presents a list for the weakest devices in terms of different strength ratio metrics.  Wind farms (GFL converters) present the lowest values for $S\scriptstyle{\rm R}_{\scriptstyle v\imath_{\rm q}}$, being the minimum ratio 6.42 for WF Knockacummer.  Regarding $S\scriptstyle{\rm R}_{\scriptstyle \gamma\imath_{\rm p}}$, synchronous units show the lowest values, with a minimum of 0.17 for TG Huntstown.  Finally, other synchronous units show the lowest $S\scriptstyle{\rm R}_{\scriptstyle \gamma\imath_{\rm p}}$ metrics.  Nevertheless, all devices face acceptable strength ratios according to the guidelines of Section \ref{subsec:sdas}.  In order to observe a deteriorated condition triggering a dynamic issue, we increase the power of WF Knockacummer so that its ratio metric $S\scriptstyle{\rm R}_{\scriptstyle v\imath_{\rm q}}$ drops below 2.00.  The wind farm performance is tested in a time-domain simulation for a three-phase short-circuit at the nearby Ballynahulla 110 kV substation.  The fault is cleared after 120 ms by opening Glenlara - Ballynahulla 110 kV cable.  Fig.  \ref{fig:voltage_device} depicts the trajectories of the terminal voltage and active power injected by WF Knockacummer for different strength ratio cases.  As expected, the device's performance worsens for lower strength ratios.  For the low ratio 1.80, the device cannot ride through the fault and the simulation collapses.

\begin{table}[hbtp]
\renewcommand{\arraystretch}{1.2}
\caption{Lowest ratio metrics devices for $S\scriptstyle{\rm R}_{\scriptstyle v\imath_{\rm q}}$, $S\scriptstyle{\rm R}_{\scriptstyle \gamma\imath_{\rm p}}$ and $S\scriptstyle{\rm R}_{\scriptstyle \sigma\imath_{\rm q}}$.}
\centering
\begin{tabular}{l|c|c|rrr}
\hline
\multicolumn{1}{c|}{Device} & Type & MVA & \multicolumn{1}{c}{$S\scriptstyle{\rm R}_{\scriptstyle v\imath_{\rm q}}$} & \multicolumn{1}{c}{$S\scriptstyle{\rm R}_{\scriptstyle \gamma\imath_{\rm p}}$} & \multicolumn{1}{c}{$S\scriptstyle{\rm R}_{\scriptstyle \sigma\imath_{\rm q}}$} \\ \hline
WF Knockacummer              & GFL  & 45  & 6.32                      & 22.48                     & 3.91                       \\
WF Barranafaddock            & GFL  & 12  & 6.48                      & 122.45                    & 14.50                      \\
WF Grousemount               & GFL  & 46  & 6.71                      & 19.97                     & 3.52                       \\
%WF Scartaglen                & GFL  & 12  & 6.85                      & 69.19                     & 14.30                      \\
TG Huntstown                 & SG   & 427 & 8.73                      & 0.17                      & 0.32                       \\
HY Ardnacrusha G2            & SG   & 25  & 14.75                     & 0.18                      & 0.68                       \\
HY Cathaleen’s Fall G4       & SG   & 26  & 14.17                     & 0.21                      & 2.31                       \\
%HY Turlough Hill G2          & SG   & 72  & 12.04                     & 0.21                      & 0.42                       \\
CCGT Great Island            & SG   & 480 & 8.47                      & 0.26                      & 0.11                       \\
CCGT Whitegate               & SG   & 445 & 8.63                      & 0.82                      & 0.12                       \\
CCGT Aghada                  & SG   & 435 & 7.85                      & 1.31                      & 0.15                       \\
%CCGT Tynagh                  & SG   & 270 & 9.60                      & 0.30                      & 0.18                       \\
\hline
\end{tabular}
\label{tab:weakdev}
\end{table}

\subsection{Strength-based hosting capacity assessment}
This subsection illustrates a practical application of the proposed framework for strength-based hosting-capacity assessment.  By combining device-level requirements with  bus-level strength metrics, the maximum admissible size of a prospective device can be estimated for different connection points.

\begin{figure}[hbtp]

    \centering
    \includegraphics[width=\linewidth, trim={0 1cm 0 0},clip]{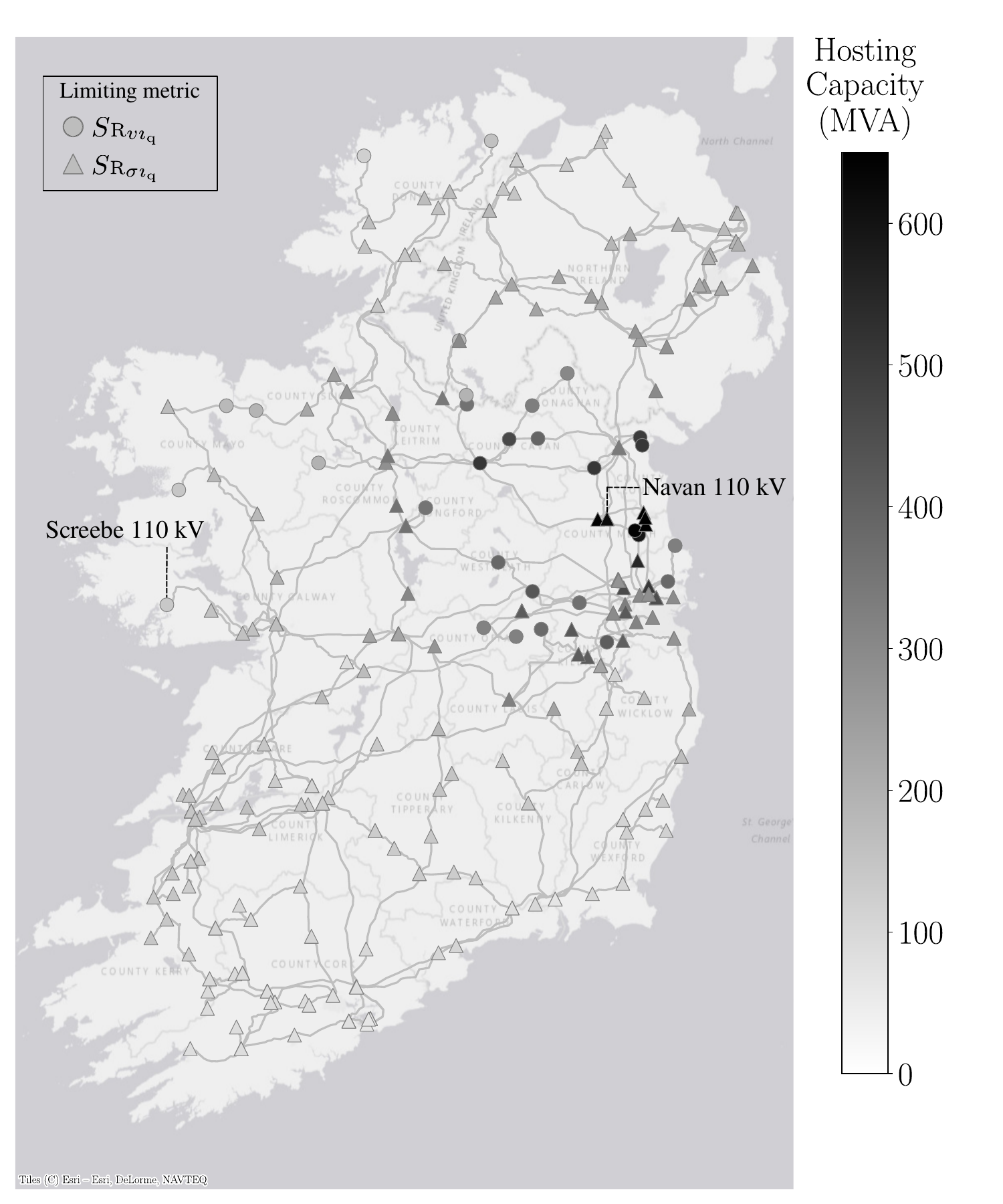}

    \caption{Hosting capacity at each HV bus for the hypothetical converter.}
    \label{fig:MAP_hosting}

\end{figure}

Consider an hypothetical prospective device that has gone through a strength assessment against the strength source.  Resulting from this analysis, the minimum strength-ratios required by the device for proper operation are identified.  For example, take a GFL converter as the one analysed in Section \ref{subsec:sdas} with the following limits: $S\scriptstyle{\rm R}_{\scriptstyle v\imath_{\rm q}}^{\rm min}$ = 3.0 and $S\scriptstyle{\rm R}_{\scriptstyle \sigma\imath_{\rm q}}^{\rm min}$ = 2.0.
Given the device limits and network strength metrics, the maximum admissible size can be calculated at every location.  Fig.  \ref{fig:MAP_hosting} illustrates a map with the hosting capacity available at every HV substation of the system for our hypothetical GFL converter.  For some substations the limiting metric was $S\scriptstyle{\rm R}_{\scriptstyle v\imath_{\rm q}}^{\rm min}$, marked with circles, and for others the limiting metric was $S\scriptstyle{\rm R}_{\scriptstyle \sigma\imath_{\rm q}}^{\rm min}$, marked with triangles.

\begin{figure}[t]

    \centering
    % Top figure
    \includegraphics[width=0.6\linewidth]{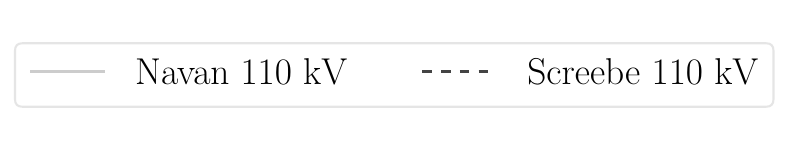}
    % Bottom left
    \begin{minipage}{0.49\linewidth}
        \centering
        \includegraphics[width=\linewidth]{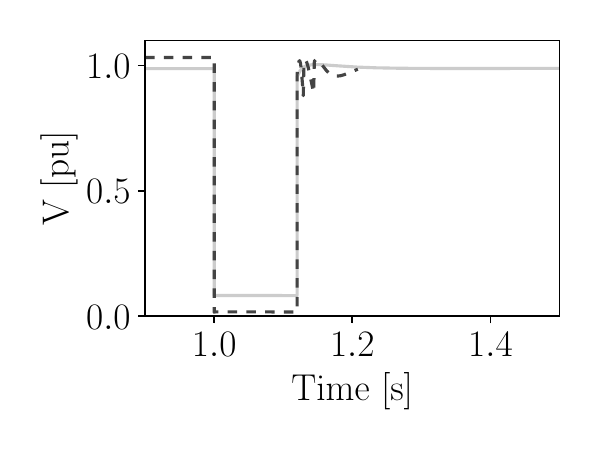}
    \end{minipage}
    \hfill
    % Bottom right
    \begin{minipage}{0.49\linewidth}
        \centering
        \includegraphics[width=\linewidth]{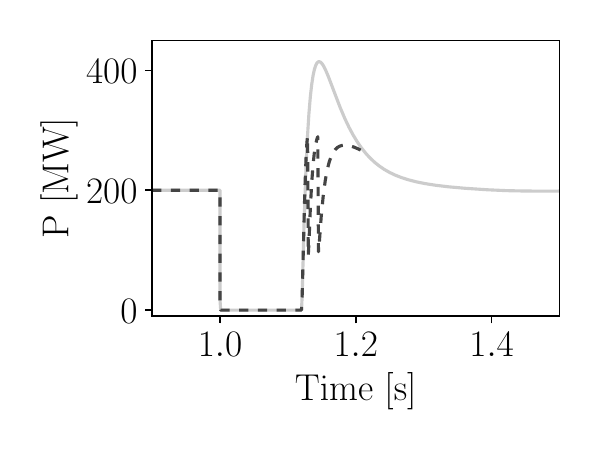}
    \end{minipage}

    \caption{Trajectories of the terminal voltage (left panel) and active power (right panel) from the hypothetical device at two different locations.}
    \label{fig:hyp}

\end{figure}

Suppose the rated power of the device is set at 200 MVA.  Two different point of connections are tested.  First, Navan 110 kV, which has vastly enough hosting capacity for the hypothetical device (640 MVA).  Second, Screebe 110 kV, whose hosting capacity is lower than the desired rating of the device (140 MVA), being $S\scriptstyle{\rm R}_{\scriptstyle v\imath_{\rm q}}{\rm min}$ the limiting metric.  Time-domain simulations for a three-phase short-circuit event at terminals are performed for both cases, and the results shown in Fig.  \ref{fig:hyp}.  The trajectories verify an inadequate operation if connected to Screebe 110 kV.  

\subsection{Remarks and guidelines}
\subsubsection{On the relationship between strength and stability}
Strength and stability are different concepts with independent definitions, but closely related in practice.  The root of their link is at their common dependence on system parameters.  For example, a change in the inertia of a synchronous machine (i) alters system strength metrics, and (ii) potentially leads to instability.  The analysis shows that weaker systems are generally more prone to instability.  Nonetheless, it is important to bear in mind that strength and stability are fundamentally distinct concepts.  Strength analysis must not be confused with a surrogate to stability analysis.  A good level of strength is a desirable property of the system by itself, less sensible to current injection changes and thus harder to disturb.  

\subsubsection{Characterizing acceptable levels of strength}

Acceptable levels of strength are device and system dependent.  A synchronous machine might tolerate a certain strength condition and another device might not.  A systematic screening for each device is an effective way to determine its strength limits.  If conducted for all devices of the system, the transmission system operator (TSO) can define acceptable operational strength limits based on the strength limits declared by each device.  In addition, the TSO can also support the definition of the acceptable levels taking into account system protection settings.  This leverages the physical domain of our metrics, which can be directly linked to voltage, frequency and RoCoF protection limits.  Specific rules, safety margins, and methodologies require further study.

\section{Conclusion}\label{sec:conclusion}
This paper provides a practical implementation of the analytical framework for power system strength developed in a prior paper.  A simplified formulation is proposed, featuring twelve individual strength metrics.  However, given the typical parameters of real bulk power systems, some metrics are either greatly correlated---such as $S_{v\imath_{\rm q}}$ and $S_{\theta\imath_{\rm p}}$---, null---first-order metrics---, or have a relatively very small value when compared to others of the same order.  Key metrics for the real case studied are $S_{v\imath_{\rm q}}$, $S_{\theta\imath_{\rm p}}$, $S_{\sigma\imath_{\rm q}}$, $S_{\gamma\imath_{\rm p}}$.  Results show that the role of first-order metrics remains less critical given the null first-order components of system devices.

The paper demonstrates in a realistic study case how bus-level strength calculation accurately characterizes the bus dynamic response after disturbances, and device-level strength calculation allows for a comprehensive screening of all devices to identify potential dynamic issues.

Future work will further explore the link between strength and stability, define benchmarks to determine the minimum device strength limits, and investigate strength-based hosting capacity assessment, as well as other applications for commissioning, security assessment and planning.

%Future work will further explore the link between strength and stability, define benchmarks for minimum strength limits, and investigate strength-based hosting-capacity assessment as well as applications to commissioning, security assessment, and planning.

%\bibliographystyle{IEEEtran}
%\bibliography{ref}


%\vspace{-6mm}

%\begin{IEEEbiography}[{\includegraphics[width=1in, height=1.25in, clip, keepaspectratio]{Images/ignacio.pdf}}] {Ignacio Ponce} received from the University of Chile the BSc.~and MSc.~degree in Electrical Engineering in 2019 and 2022, respectively.  He is currently pursuing a Ph.D in Electrical Engineering at University College Dublin, Ireland.  His research interests include power system modeling, control and stability analysis.
%\end{IEEEbiography}

%\vspace{-6mm}

%\begin{IEEEbiography}[{\includegraphics[width=1in, height=1.25in, clip, keepaspectratio]{Images/federico.pdf}}] {Federico Milano} (F'16) received from the Univ.~of Genoa, Italy, the Ph.D.~in Electrical Engineering 2003.  In 2013, he joined the University College Dublin, Ireland, where he is Professor of Power System Dynamics.  He is Chair of the IEEE Power System Stability Controls Subcommittee, IET Fellow, IEEE PES Distinguished Lecturer, Editor of the IEEE Power \& Energy Magazine, and Co-Editor in Chief of the IET Generation, Transmission \& Distribution.  His research interests include power system modeling, control and stability analysis.
%\end{IEEEbiography}

\vfill

\onecolumn
\appendix[Strength components of devices]
% Synchronous machines
\renewcommand{\azero}{
{\setlength{\arraycolsep}{2pt}
\begin{bmatrix}
r_{\rm a} & -x_{\rm q}'\\
x_{\rm d}' & r_{\rm a}
\end{bmatrix}
}
    }
\renewcommand{\aone}{0}
\renewcommand{\bone}{\azero}

\renewcommand{\atwo}{\frac{\Omega_{\rm b}}{M}\ull{\rm J}\,\ull{a}\, \ull{E}\, \M{1}{0}{0}{0}\, \ull{E}^{*}\, \ull{a}}
\renewcommand{\btwo}{0}
\renewcommand{\ctwo}{\azero}

\begin{table*}[hbtp]

\begin{minipage}{0.5\linewidth}
\caption{Synchronous machines' strength components.}
\centering
\begin{tabular}{l|l|l|l} 
Order&a&b&c\\
\hline
\rule{0pt}{4ex}  
Zero & $\azero$ & $0$ & $0$\\
\rule{0pt}{4ex} 
First & $\aone$ &$\bone$ & $0$\\
\rule{0pt}{4ex} 
Second & $\atwo$&$\btwo$ &$\ctwo$ \\
& & &\\
\hline
\end{tabular}
\label{tab:syn}
\end{minipage}
\hfill
\begin{minipage}{0.4\linewidth}
\centering
\vspace{3em}
\raggedright
\hspace{0.5em}Where:\\[1em]
\renewcommand{\arraystretch}{1.2}
\begin{tabular}{ll}
$r_{\rm a}$                     & armature resistance.                      \\
$x'_{\rm d}$, $x'_{\rm q}$      & dq-axis transient reactances         \\
$M$                            & mechanical starting time (=2H).          \\
$\Omega_{\rm b}$               & Base frequency.      \\
$\ull{\rm J}=\M{0}{-1}{1}{0}$  & $\jmath$ in matrix form.  \\[1em]
$\ull{E}=e'_{\rm q}\,\ull{e}^{\jmath\,\delta}$ & internal emf in matrix form.
\end{tabular}
\end{minipage}

\end{table*}

% Constant Z Load
\renewcommand{\azero}{
\setlength{\arraycolsep}{2pt}
-\begin{bmatrix}
    r_{\rm l} & -x_{\rm l}\\
    x_{\rm l} & r_{\rm l}
    \end{bmatrix}^{-1}
    }
\renewcommand{\aone}{0}
\renewcommand{\bone}{\azero}

\renewcommand{\atwo}{0}
\renewcommand{\btwo}{0}
\renewcommand{\ctwo}{\azero}

\begin{table*}[hbtp]

\begin{minipage}{0.5\linewidth}
\caption{Constant impedance loads' strength components.}
\centering
\begin{tabular}{l|l|l|l} 
Order&a&b&c\\
\hline
\rule{0pt}{4ex}  
Zero & $\azero$ & $0$ & $0$\\
\rule{0pt}{4ex} 
First & $\aone$ &$\bone$ & $0$\\
\rule{0pt}{4ex} 
Second & $\atwo$&$\btwo$ &$\ctwo$ \\
& & &\\
\hline
\end{tabular}
\label{tab:zload}
\end{minipage}
\hfill
\begin{minipage}{0.4\linewidth}
\centering
\vspace{3em}
\raggedright

\hspace{0.5em}Where:\\[1em]

\renewcommand{\arraystretch}{1.25}
\begin{tabular}{ll}
$r_{\rm l}$ & load resistance. \\
$x_{\rm l}$ & load reactance.
\end{tabular}

\end{minipage}

\end{table*}

% GFL IBR
\renewcommand{\azero}{
\ull{\imath}\M{-1}{0}{0}{1}\ull{v}^{-1}-\ull{y_{\rm s}}
    }
\renewcommand{\aone}{0}
\renewcommand{\bone}{\ull{a}+\ull{v^{*}}^{-1}\M{-T_{\rm g}}{0}{0}{T_{\rm g}}^{-1}\ull{K_{\rm p}}\,\ull{D}\,\ull{v}^{-1}}

\renewcommand{\atwo}{0}
\renewcommand{\btwo}{\ull{v^{*}}^{-1}\left(\frac{1}{T_{\rm g}^2}(\ull{K_{\rm p}}^2+\ull{K_{\rm p}})-\frac{1}{T_{\rm g}}\ull{K_{\rm i}}\right)\,\ull{D}\,\ull{v}^{-1}}
\renewcommand{\ctwo}{\ull{b}'}

\begin{table*}[hbtp]

\begin{minipage}{0.5\linewidth}
\caption{Grid following IBRs' strength components.}
\centering
\begin{tabular}{l|l|l|l} 
Order&a&b&c\\
\hline
\rule{0pt}{4ex}  
Zero & $\azero$ & $0$ & $0$\\
\rule{0pt}{4ex} 
First & $\aone$ &$\bone$ & $0$\\
\rule{0pt}{4ex} 
Second & $\atwo$&$\btwo$ &$\ctwo$ \\
& & &\\
\hline
\end{tabular}
\label{tab:gfl}
\end{minipage}
\hfill
\begin{minipage}{0.4\linewidth}
\centering
\vspace{3em}
\raggedright
\hspace{0.5em}Where:\\[1em]

\renewcommand{\arraystretch}{1.25}
\begin{tabular}{ll}
$T_{\rm g}$ & P control lag time constant. \\
$\ull{y_{\rm s}}=\M{g_{\rm s}}{-b_{\rm s}}{b_{\rm s}}{g_{\rm s}}$ & \begin{tabular}{@{}l@{}}output shunt admittance\\ in matrix form.\end{tabular}\\[1em]
$\ull{D}=\M{0}{D}{0}{0}$ & inverse droop in matrix form.\\[1em]
$\ull{K_{\rm p}}=\M{K_{\rm p}^{\rm p}}{0}{0}{K_{\rm p}^{\rm q}}$ & \begin{tabular}{@{}l@{}}PQ control proportional\\ gains in matrix form.\end{tabular}\\[1em]
$\ull{K_{\rm i}}=\M{K_{\rm i}^{\rm p}}{0}{0}{K_{\rm i}^{\rm q}}$ & \begin{tabular}{@{}l@{}}PQ control integral\\ gains in matrix form.\end{tabular}
\end{tabular}

\end{minipage}

\end{table*}

\vfill

\end{document}